\documentclass[aps,prx,twocolumn,superscriptaddress,floatfix,amssymb]{revtex4-2}

\usepackage{graphicx,dcolumn,bm,xstring,braket,amsmath,comment}
\usepackage[unicode=true,pdfusetitle,bookmarks=true,bookmarksnumbered=true,bookmarksopen=false, breaklinks=false,pdfborder={0 0 0},pdfborderstyle={},backref=false,colorlinks=false, pdftitle={Absence of heating in a uniform Fermi gas created by periodic driving}]{hyperref}

\newcommand{\cog}[1]{\left({#1}\right)}

\newcommand{\psid}[2]{
    \IfEqCase{#1}{
        {1}{\hat{\Psi}^\dagger_\uparrow(\mathbf{#2})}
        {2}{\hat{\Psi}^\dagger_\downarrow(\mathbf{#2})}
        {s}{\hat{\Psi}^\dagger_s(\mathbf{#2})}
        {s'}{\hat{\Psi}^\dagger_{s'}(\mathbf{#2})}
    }[\PackageError{psid}{Undefined option to psid: #1}{}]
}

\newcommand{\psind}[2]{
    \IfEqCase{#1}{
        {1}{\hat{\Psi}_\uparrow(\mathbf{#2})}
        {2}{\hat{\Psi}_\downarrow(\mathbf{#2})}
        {s}{\hat{\Psi}_s(\mathbf{#2})}
        {s'}{\hat{\Psi}_{s'}(\mathbf{#2})}
    }[\PackageError{psi}{Undefined option to psi: #1}{}]
}

\begin{document}

\title{Absence of heating in a uniform Fermi gas created by periodic driving}

\author{Constantine Shkedrov}
\affiliation{Physics Department and Solid State Institute, Technion - Israel Institute of Technology, Haifa 32000, Israel}
\author{Meny Menashes}
\affiliation{Physics Department and Solid State Institute, Technion - Israel Institute of Technology, Haifa 32000, Israel}
\author{Gal Ness}
\affiliation{Physics Department and Solid State Institute, Technion - Israel Institute of Technology, Haifa 32000, Israel}
\author{Anastasiya Vainbaum}
\affiliation{Physics Department and Solid State Institute, Technion - Israel Institute of Technology, Haifa 32000, Israel}
\author{Ehud Altman}
\affiliation{Department of Physics, University of California, Berkeley, California 94720, USA}
\affiliation{Materials Sciences Division, Lawrence Berkeley National Laboratory, Berkeley, California 94720, USA}
\author{Yoav Sagi}
\email[Electronic address: ]{yoavsagi@technion.ac.il}
\affiliation{Physics Department and Solid State Institute, Technion - Israel Institute of Technology, Haifa 32000, Israel}

\date{\today}

\begin{abstract}
Ultracold atomic gas provides a useful tool to explore many-body physics. One of the recent additions to this experimental toolbox is the Floquet engineering, where periodic modulation of the Hamiltonian allows the creation of effective potentials that do not exist otherwise. When subject to external modulations, however, generic interacting many-body systems absorb energy, thus posing a heating problem that may impair the usefulness of this method. For discrete systems with bounded local energy, an exponentially suppressed heating rate with the driving frequency has been observed previously, leaving the system in a prethermal state for exceedingly long durations. But for systems in continuous space, the situation remains unclear. Here we show that Floquet engineering can be employed to a strongly interacting degenerate Fermi gas held in a flat box-like potential without inducing excessive heating on experimentally relevant timescales. The driving eliminates the effect of a spin-dependent potential originating from a simultaneous magnetic levitation of two different spin states. We calculate the heating rate and obtain a power-law suppression with the drive frequency. To further test the many-body behavior of the driven gas, we measure both the pair-condensation fraction at unitarity and the contact parameter across the BEC-BCS crossover. At low driving frequencies, the condensate fraction is reduced by the time-dependent force, but at higher frequencies, it revives and attains an even higher value than without driving. Our results are promising for future exploration of exotic many-body phases of a bulk strongly-interacting Fermi gas with dynamically engineered Hamiltonians.
\end{abstract}

\maketitle

\section{Introduction}

The last two decades have witnessed tremendous advance in studying many-body problems with ultracold atomic gases \cite{RevModPhys.80.885}. The vast majority of works have been done with static Hamiltonians. Adding periodic driving can generate effective Hamiltonians with completely different properties than the original one -- an approach called Floquet engineering \cite{Goldman2014,Bukov2015,Eckardt2017}. For example, modulation of the barrier between potential wells renormalizes the tunneling rate \cite{Grossmann1991,Grifoni1998,Lignier2007,Kierig2008} and can drive quantum phase transitions \cite{Zenesini2009}. Floquet engineering can also be used to create artificial gauge fields \cite{Soerensen2005,Aidelsburger2013,Miyake2013,Jotzu_2014}, and give rise to new phases which do not exist at equilibrium \cite{Rudner2020,Wintersperger2020}. 

An inherent problem with externally driven systems is their tendency to heat up. Apart from integrable and many-body localized systems, generic interacting ensembles will absorb energy from the external force and eventually reach an ``infinite temperature'' where all states are equally populated \cite{Lazarides2014,DAlessio2014,Ponte2015}. Nonetheless, recent theoretical works suggest that in discrete lattice systems, the energy absorption rate is generally exponentially small in the drive frequency over the energy of a local excitation  \cite{Mori2016,Kuwahara2016,Abanin2017,Else2017,Machado2019,PhysRevLett.125.080602}. These predictions are supported by heating rate measurements done with bosons in a driven optical lattice \cite{PhysRevX.10.021044}. The exponential suppression of heating in discrete systems relies on the fact that the energy is locally bounded. However, in continuous systems (e.g., bulk quantum gases), there is no such bound. The pertinent question, in this case, is under what conditions one can obtain ``cold'' prethermal states exhibiting collective phenomena that are governed by an effective ``Floquet engineered'' Hamiltonian. 

In this paper we address this question using a driven ultracold Fermi gas near unitarity, exhibiting high-$T_c$ fermionic superfluidity \cite{Zwerger2012}. The driving we apply is intended to create a uniform effective potential across the Fermi gas. Although \textit{in situ} measurements \cite{Schirotzek2008,Nascimbene2010,Horikoshi2010,Ku03022012} and spatial selection \cite{Miller2007,Drake2012,Sagi2012,Sagi2013,Sagi2015,Carcy2019} can give access to quasi-homogeneous observables, it is better to create a uniform gas from the outset.
This is essential, for example, to study critical properties and avoid spurious phase separated states \cite{Shin2006}.
Indeed, in recent years, uniform Bose \cite{Gaunt2013,Chomaz2015} and Fermi \cite{Mukherjee2017,Hueck2018} gases have been created in flat optical traps. These traps are formed by several shaped laser beams that create sharp repelling walls enclosing a dark volume. 

\begin{figure}[t]
	\centering
	\includegraphics[width=1\linewidth]{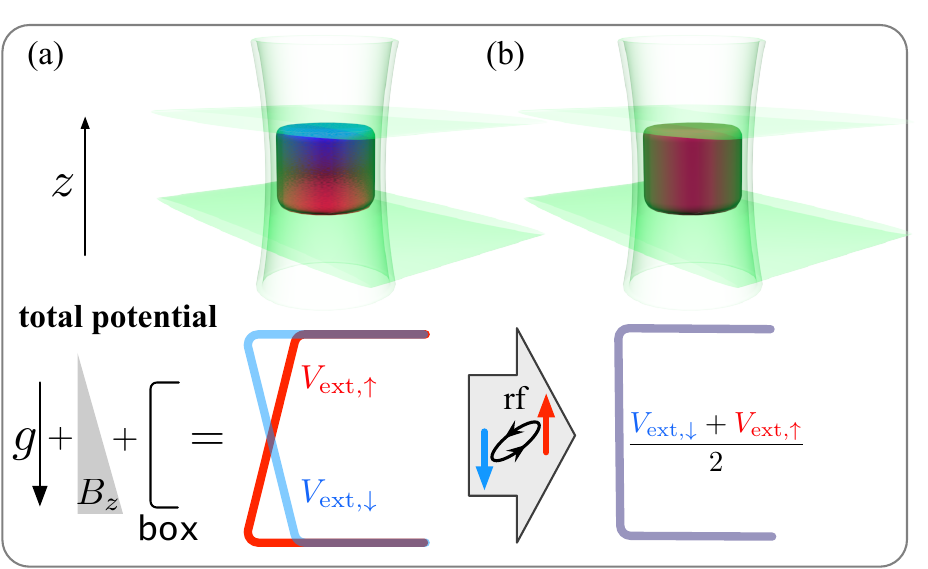}
	\caption{\textbf{Creating a uniform Fermi gas by periodic driving.} \textbf{(a)} The gas, composed of two spin states (marked by red and blue colors and by opposite arrows), is trapped in a box-like optical potential. The two spins have different magnetic dipole moments. As a result, it is only possible to partially counteract the gravitational potential with a magnetic field gradient, $B_z$, set according to Eq.(\ref{Eq:average_magnetic_gradient}). The total external potential V$_{\textrm{ext},s}$ depends on the spin, $s\in\{\uparrow,\downarrow\}$, and consequently, the density distribution of each spin is different and not uniform (color gradient in the top left figure). \textbf{(b)} By adding a resonant rf field that drives rapid spin rotations, we create an effective spin-independent potential, in which the gas becomes homogeneous. Importantly, the intrinsic many-body behavior of the gas is unchanged by this driving.}
	\label{fig:sketch}
\end{figure}

A significant challenge is posed by the need to offset the gravitational potential, which leads to a substantial energy change in the trap. One obvious solution is to use a shaped optical potential to counter gravity \cite{Shibata2020}. But generating such a potential, smooth on a nano-Kelvin scale, is a formidable task. A simpler approach taken in previous experiments is to use a magnetic field with an appropriate gradient. This works if all particles have approximately the same magnetic dipole moment. However our $^{40}$K Fermi gas is a mixture of two hyperfine states with different (but not opposite) magnetic moments $\mu_\uparrow\ne\mu_\downarrow$. An appropriate magnetic field gradient counters the average gravitational potential, while leaving an opposite potential gradient on each of the two species (Fig. \ref{fig:sketch}a). 

To counter the residual field gradient we apply an rf field that induces a rapid percession at a Rabi frequency $\Omega$. In the rotating frame of the percessing spins, the static spin dependent potential gradient is translated to a periodic perturbation of frequency $\Omega$. The rest of the interacting fermion Hamiltonian, including the flat spin-independent potential, is invariant to spin rotations and therefore unchanged in the rotating frame (Fig. \ref{fig:sketch}b). Thus we achieve an ultracold uniform state of a spin-balanced gas of $^{40}$K atoms that is useful if the periodic perturbation does not cause significant heating over experimental time scales.

We establish this property by measuring the pair-condensate fraction (CF) at unitarity while applying a continuous driving. At low frequencies, driving impairs the gas conditions and reduces the CF. As the frequency increases further, the CF recovers and even surpasses its value without the driving. At high driving frequencies, we do not detect heating or excessive loss of atoms which can be attributed to the drive. Finally, we perform rf spectroscopy with a uniform gas in the BEC-BCS crossover regime, and extract the homogeneous contact parameter as a function of the interaction strength.

The structure of this paper is as follows. In section \ref{Theory_sec} we review the theoretical model for radio frequency driving and calculate the expected heating rate in presence of this drive. In section \ref{Experiment_sec}, we describe the experimental setup and measurement sequence. The results are presented in section \ref{Results_sec}. We study with time-dependent \emph{in situ} imaging the relaxation dynamics following the application of the driving field. The temperature of the uniform gas is probed by Raman spectroscopy. The many-body behavior of the uniform driven gas is studied with a pair-projection technique and rf spectroscopy. We study both the frequency dependence and the long time behavior of the driven gas. Section \ref{Discussion_sec} concludes with a discussion and outlook.

\section{Theoretical model}\label{Theory_sec}

\textbf{\textbf{Effective Hamiltonian.}---} We consider fermions in two possible spin states, denoted by $\left|\downarrow\right\rangle$ and $\left|\uparrow\right\rangle$, where the energy of the latter is larger by $\hbar \omega_0$. The two particles are placed in an external potential and coupled by an rf field with a frequency $\omega_\mathrm{rf}$. The Hamiltonian is a sum of three terms $\hat{H}=\hat{H}_0+\hat{H}_\mathrm{int}+\hat{H}_\mathrm{rf}$ that account for the single-particle kinetic and potential energy ($\hat{H}_0$), the interaction Hamiltonian ($\hat{H}_\mathrm{int}$) and the coupling to the external rf field ($\hat{H}_\mathrm{rf}$). In the frame rotating with the $\uparrow$ spin, they are given by \cite{giorgini2008theory,Toermae2014}
\begin{subequations}\label{eq_Hamiltonian}
\begin{equation}\label{eq_Hamiltonian_1a}
    \hat{H}_0= 
    \sum_{\sigma=\uparrow,\downarrow}\int d^3 r \hat{\Psi}^{\dagger}_{\sigma}(\mathbf{r})\left(-\frac{\hbar^2\nabla^2}{2m}+V_\sigma(\mathbf{r})\right)\hat{\Psi}_{\sigma}(\mathbf{r})
\end{equation}
\begin{equation}\label{eq_Hamiltonian_1b}
    \hat{H}_\mathrm{int}= \sum_{\sigma\sigma'}\iint d^3 r' d^3 r u(\mathbf{r}-\mathbf{r'})\hat{\Psi}^\dagger_\sigma(\mathbf{r})\hat{\Psi}^\dagger_{\sigma'}(\mathbf{r'})\hat{\Psi}_{\sigma'}(\mathbf{r'})\hat{\Psi}_\sigma(\mathbf{r})
\end{equation}
\begin{equation}\label{eq_Hamiltonian_1c}
    \hat{H}_\mathrm{rf}
    = \frac{\hbar}{2}\int d^3 r\  \Omega e^{i\omega_0 t}\cog{e^{i\omega_{\textbf{rf}}t}+e^{-i\omega_{\textbf{rf}}t}}\hat{\Psi}^{\dagger}_{\uparrow}(\mathbf{r})\hat{\Psi}_{\downarrow}(\mathbf{r})+\mathrm{h.c.} \, \, ,
\end{equation}
\end{subequations}
where $\Omega$ is the Rabi frequency of the rf field, $V_{\sigma}(\mathbf{r})$ is the external potential for spin $\sigma$, and $\hat{\Psi}_\sigma(\mathbf{r})$ are fermionic field operators obeying the anti-commutation relation $\{{\hat{\Psi}_\sigma(\mathbf{r})},{\hat{\Psi}^\dagger_{\sigma'}(\mathbf{r'})}\}=\delta_{\sigma\sigma'} \delta(\mathbf{r}-\mathbf{r'})$. Note that here we consider a general spin symmetric (and translationally invariant) two body interaction. The interaction potential $u(r-r')$ represents the microscopic interaction, rather than a low energy limit of the $T$-matrix (pseudopotential). In particular we will be interested in a unitary gas for which the $T$-matrix is explicitly energy dependent in the low energy limit.

In our experiment, the rf field is resonant with the bare energy difference $\omega_\mathrm{rf}=\omega_0$, and $\omega_0\gg \Omega$. Thus, within the rotating wave approximation, the rf field is seen as a large magnetic field along $S^x$, $\hat{H}_\mathrm{rf}=\frac{\hbar}{2}\int d^3 r\Omega\hat{\Psi}^{\dagger}_{\uparrow}(\mathbf{r})\hat{\Psi}_{\downarrow}(\mathbf{r})+\mathrm{h.c.}$. 
The initial state is presumed to be an ultracold Fermi gas with equally populated spin components $N_\uparrow=N_\downarrow$, which in presence of the field precesses around the $x$ axis at the Rabi frequency $\Omega$. To clearly see the effect of the external rf field, we eliminate the field by a unitary transformation, $\hat{U}=e^{\frac{i}{\hbar}\hat{H}_\mathrm{rf} t}$, into a reference frame that rotates with the spins.

The Hamiltonian transforms in a simple way under $\hat{U}$. The kinetic energy and interaction Hamiltonians are both invariant under spin rotations and are therefore unchanged by the time-dependent transformation. As first noted by Zwierlein \textit{et al.} \cite{Zwierlein_2003}, the invariance of contact interactions under rf rotations is the reason for the absence of a spectroscopic shift in the transition frequency between the spins.
The external potential, on the other hand, can be decomposed into spin symmetric and anti-symmetric parts
\begin{align}
\sum_{\sigma=\uparrow,\downarrow} V_\sigma(\mathbf{r})\hat{n}_\sigma(\mathbf{r})=V(\mathbf{r})\hat{n}(\mathbf{r})-h(\mathbf{r})\hat{S}^z(\mathbf{r}) \ \ ,
\end{align}
where $\hat{n}(\mathbf{r})=\hat{n}_\uparrow(\mathbf{r})+\hat{n}_\downarrow(\mathbf{r})$
is the number density, $\hat{S}^z(\mathbf{r})=(\hat{n}_\uparrow(\mathbf{r})-\hat{n}_\downarrow(\mathbf{r}))/2$ the spin density, $V(\mathbf{r})=(V_\uparrow(\mathbf{r})+V_\downarrow(\mathbf{r}))/2$ and $h(\mathbf{r})=V_\downarrow(\mathbf{r})-V_\uparrow(\mathbf{r})$. 
Only the spin symmetric part of the potential, which couples to the number density, is invariant under the transformation $\hat{U}$ and does not change with time.

In our experiment, the external potential is given by $V_\sigma(\mathbf{r})=V_{\mathrm{trap}}(\mathbf{r})+mgz-\mu_\sigma B' z$, where the first term is the flat optical potential, the second term is the gravitational potential, and the last term describes the interaction of a spin with a magnetic moment $\mu_\sigma$ with the external magnetic field, which is linear in height $z$. Thus, the spin symmetric part of the potential is $V(\mathbf{r})=V_{\mathrm{trap}}(\mathbf{r})+mgz-B' z(\mu_\uparrow+\mu_\downarrow)/2$.
By tuning the magnetic field gradient to the value
\begin{equation}
\label{Eq:average_magnetic_gradient}
    B'=\frac{2mg}{\mu_\uparrow+\mu_\downarrow} \ \ 
\end{equation} 
we obtain a flat total potential. The gravitational potential is thus eliminated by the static part of the Hamiltonian. 

The spin anti-symmetric part of the external potential, which couples to the spin density, gives rise to a time-dependent coupling in the rotating frame $\hat{U}$. Specifically, the spin density rotates as $\hat{U}\hat{S}^z\hat{U}^\dagger= \cos(\Omega t) \hat{S}^z -\sin(\Omega t)\hat{S}^y$, leading to a Zeeman field rotating at frequency $\Omega$ in the $\hat{S}^z, \hat{S}^y$ plane. With a simple redefinition of the spin axes we write the residual time-dependent part of the Hamiltonian as a rotating field in the $\hat{S}^x,\hat{S}^y$ plane 
\begin{align}
    \hat{V}_a(t)=-\frac{V_g}{L} \int d^3 r \,z \,\theta\left(L/2-|z|\right) \left( e^{-i\Omega t} \hat{S}^+(\mathbf{r})+ \mathrm{h.c.}\right) \ \ .
    \label{eq:tdep}
\end{align}
Here $\hat{S}^+(\mathbf{r})=\hat{\Psi}^\dagger_\uparrow(r)\hat{\Psi}_\downarrow(r)$, $\theta(z)$ is the Heaviside step function, and $L$ is the height of the box potential. $V_g=mgL(\mu_\uparrow-\mu_\downarrow)/(\mu_\uparrow+\mu_\downarrow)$, where we use Eq.(\ref{Eq:average_magnetic_gradient}) to tune a flat static potential.   

\textbf{\textbf{Heating rate.}---} The heating rate due to irreversible transitions induced by the residual time-dependent term (\ref{eq:tdep}) in the effective potential can be calculated from the linear response of the system to this perturbation. 
For this purpose it is convenient to rewrite the time-dependent perturbation in momentum space
\begin{align}
    \hat{V}_a(t)=\sum_q v_q e^{-i\Omega t}\hat{S}^+_q + \mathrm{h.c.} \ \ ,
\end{align}
where $\hat{S}_q^+=\int d^3 r \hat{S}^+(\mathbf{r})e^{-i \mathbf{q}\cdot \mathbf{r}}$ and $v_q$ is the spatial Fourier transform of the time-dependent effective potential 
\begin{align}
    v_q=-i V_g \delta_{q_x,0}\delta_{q_y,0} \frac{L q\cos\left(L q/2\right)-2\sin\left(L q/2\right)}{(L q)^2}e^{-(\lambda_0 q)^2}.
    \label{eq:vq}
\end{align}
The Gaussian fall off at large $q$ is due to convolution with the optical resolution that limits the sharpness of the potential features. In our experiment the resolution $\lambda_0\approx 3 \mathrm{\mu}$m happens to be close to the Fermi wavelength $\lambda_F\approx 2.6 \mathrm{\mu}$m, so for simplicity we shall identify the two scales and use $\lambda_F$ also as the resolution limit.

\begin{figure}[t]
	\centering
	\includegraphics{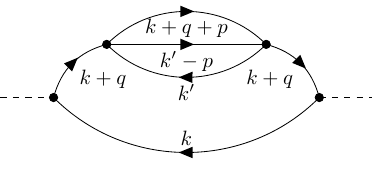}
	\caption{\textbf{Linear response to the residual time-dependent potential.}
		In the diagram, the incoming and outgoing dashed lines represent the operators $\hat{S}_q^+$ and $\hat{S}_q^-$ attached at the vertices with the couplings $V_q$ and $V_{-q}$ respectively. The vertices connecting four solid lines represent the contact $s$-wave interactions.}
	\label{fig:diagram}
\end{figure}

The transition rate $\Gamma_q$ induced by the perturbation at wave-vector $q$ is directly related to the spectral function of the operator $\hat{S}^+_q$ at the Rabi frequency through 
\begin{align}
\Gamma_q=|v_q|^2 \text{Im}\langle \hat{S}^+_q(\Omega) \hat{S}^-_{-q}(-\Omega)\rangle_{\text{Ret}} \ \ .
\end{align}

Without interactions the transition rate is exponentially small in the high frequency $\Omega$ because the only way the perturbation can excite a resonant transition is to give the particle extremely high momentum $\hbar q\approx \sqrt{ m\hbar\Omega}$, which is suppressed by the Gaussian resolution limit in the matrix element (\ref{eq:vq}).
In presence of interactions, on the other hand, a transition can satisfy energy conservation even for small momentum-transfer by utilizing the interaction term to create another particle-hole pair. The diagram of the lowest order process that can lead to a transition is shown in Fig. \ref{fig:diagram}. Pictorially, the perturbation creates a particle hole pair with small momentum $q$ (and concomitantly low energy). The excited particle with momentum $k+q$ then scatters on another particle with momentum $k'$ inside the Fermi sea in a collision with large momentum transfer $p\gg k_F$ ($k_F$ is the Fermi wave-vector) so that the pair of particles emerging from the collision are near the required final energy $\hbar\Omega$.

The transition rate in this process can be framed as a Fermi golden rule calculation, equivalent to the imaginary part of the diagram in Fig. \ref{fig:diagram}: 
\begin{equation}
\Gamma_q={1\over \hbar}\sum_f|\langle f|\hat{T}|i\rangle|^2\delta(E_f-E_i-\hbar\Omega) \ \ ,
\end{equation}
where 
\begin{equation}
\hat{T}=\sum_k \frac{\hat{H}_{\text{int}}|k,q\rangle\langle k,q|v_q\hat{S}^+_q}{\hbar\Omega-E_n} \ \ .
\end{equation}
Here the intermediate states are the particle hole states labeled by the hole momenta $k$
\begin{equation}
    |k,q\rangle=\hat{c}^\dagger_{\uparrow,k+q}\hat{c}_{\downarrow,k}|\psi_{FS}\rangle \ \ .
\end{equation}
and the accessible final states consist of two particles and two holes with $S^z=\pm1$, such as
\begin{equation}
|k,k',p,q,s\rangle=\hat{c}^\dagger_{s,k+q+p}\hat{c}^\dagger_{-s,k'-p}\hat{c}_{\downarrow,k'}\hat{c}_{\downarrow,k}|\psi_{FS}\rangle \ \ .
\end{equation}

The effective interaction vertex merits a brief discussion. By a slight abuse of notation we write the contact interaction 
\begin{equation}
\label{eq:Hint}
\hat{H}_{\text{int}}= \frac{u}{\mathcal{V}} \sum_{kk'p} \hat{c}^\dagger_{\uparrow,k+p}\hat{c}^\dagger_{\downarrow,k'-p}\hat{c}_{\downarrow,k'}\hat{c}_{\uparrow,k} \ \ , 
\end{equation}
using the same symbol used for the interaction Hamiltonian (\ref{eq_Hamiltonian_1b}), where $\mathcal{V}$ is the volume. However, it should be noted that the effective contact interaction is really the low energy limit of the  $T$-matrix. In most cases it is given by the pseudopotential  $u=4\pi \hbar^2 a/m$ with $a$ the $s$-wave scattering length. But at unitarity the  dependence on the energy of the scattered states is important as it cuts off the divergence at any non vanishing energy. The $T$-matrix appropriate for the vertex in Fig. \ref{fig:diagram} from the Lippman-Schwinger equation at unitarity is
\begin{equation}
  u= \frac{2}{\rho(\Omega/2)}=\frac{8}{3n}\frac{\epsilon_F^{3/2}}{\sqrt{\hbar\Omega/2}} \ \ ,
\end{equation}
where $n$ in the particle density and $\rho(\epsilon)$ the single particle density of states per unit volume (DOS).
We also define an interaction energy scale $U=u n$.

Substituting into the Fermi golden rule expression we obtain:
\begin{widetext}
\begin{equation}
\Gamma_q\approx {1\over \hbar}\sum_{k,k',p,s} \left(\frac{|v_q| U}{\hbar\omega+\xi_k-\xi_{k+q}}\right)^2 n_k n_{k'} (1-n_{k+q})(1-n_{k+q+p})(1-n_{k'-p}) \delta(\xi_{k+q+p}+\xi_{k'-p}-\xi_k-\xi_{k'}-\hbar\omega) \ \ .
\end{equation}
\end{widetext}
The factors $n_k(1-n_{k+q})$ constrain the sum over $k$ to a fraction $\sim |q|/k_F$ of the Fermi surface volume. Physically, this is the phase space available for creating the low momentum particle-hole pair intermediate state. The other constraints are obeyed automatically due to the required large momentum transfer $p$ in the collision. In addition, because we assume $\hbar\Omega\gg\epsilon_F$  we shall neglect the small particle-hole energy $\xi_k-\xi_{k+q}$ in the denominator and the hole energies $-\xi_k$  and $-\xi_{k'}$ in the $\delta$-function. Thus we obtain
\begin{equation}
\Gamma_q\approx \frac{|v_q|^2 U^2}{4\hbar(\hbar\Omega)^2}\frac{|q|}{k_F}\mathcal{V}\rho(\hbar\Omega)={3N\over 16\hbar}\frac{|v_q|^2 U^2}{(\hbar\Omega\,\epsilon_F)^{3/2}}{|q|\over k_F} \ \ ,
\end{equation}
where $N$ is the particle number. The total transition rate is the integral over $\mathbf{q}=q\hat{\mathbf{z}}$ up to the resolution cutoff $q_0\approx k_F$
\begin{align}
{\Gamma}&\approx {N\over\hbar}\frac{3}{64\pi}\frac{V_g^2 U^2}{(\hbar\Omega\,\epsilon_F)^{3/2}}\left({1 \over k_F L}\right)\ln\left({k_F L\over \pi}\right) \ \ .
\end{align}
The factor $(k_F L)^{-1}$ stems from the suppressed phase space for excitation of the fermi sea with low momentum transfer, while the logarithmic correction is due to the integration over the $1/q$ behavior of the function $|v_q|^2 q$ for $q\gg\pi/L$.

We remark that the scaling with $\Omega$ can be understood without a calculation by noting that the transition rate is a second order process with the intermediate states constrained to low energy compared to $\Omega$ due to the momentum cutoff on the perturbation. Thus the energy denominator due to the virtual transition is $\sim\Omega$ and the transition rate must scale as $\Gamma\sim \left({V_g U\over \Omega}\right)^2\rho(\Omega)$. Plugging in the DOS in 3d we get the correct scaling with $\Omega$. The small momentum transfer $q$ limits the phase space for intermediate states giving the $|q|/k_F$ factor in $\Gamma_q$.

Finally we can convert the transition rate to a heating rate by multiplying it with the average energy increase per particle per unit time. In our experiment, the energy $\hbar\Omega/2$ given to the excited particle pair is sufficient to escape the trap. Therefore, the heating is due to the holes left in the Fermi sea, leading to the heating rate $\dot\epsilon=\epsilon_F\Gamma/N$. 
To get a dimensionless measure of the heating rate we define $\eta=\hbar\dot{\epsilon}/\epsilon_F^2$. $\eta$ gives the relative energy change per particle $\delta \epsilon /\epsilon_F$, which occurs in the characteristic timescale of the system $\tau_F=\hbar/\epsilon_F$. If $\eta\ll 1$, we can observe phenomena slow on the scale of $\epsilon_F$ without being affected by the heating. At unitarity we get
\begin{equation}\label{eq_eta}
    \eta={2\over 3\pi}\left(\frac{V_g}{\hbar\Omega}\right)^2\sqrt{\frac{\epsilon_F}{\hbar\Omega}}\,{1\over k_F L}\ln\left(\frac{k_FL}{\pi}\right) \ \ .
\end{equation}
In our experiment, we can reach driving frequencies $\Omega/2\pi>10$kHz where this parameter is $\eta<10^{-4}$, so heating is not expected to be a problem. Before proceeding we note that a semi-classical calculation of the heating induced by periodic forces acting on an ultracold gas was reported in Ref. \cite{PhysRevA.100.033406}.

\section{Experiment}\label{Experiment_sec} 
Our experiments are performed with a quantum degenerate gas of $^{40}\mathrm{K}$ atoms, prepared in an incoherent, spin-balanced mixture of the two lowest energy states, $\left|\downarrow\right\rangle=|9/2,-9/2\rangle$ and $\left|\uparrow\right\rangle=|9/2,-7/2\rangle$, with the notation $|F,m_F\rangle$. The flat trap, $V_{\mathrm{trap}}$, is created by three laser beams with a wavelength of $532\mathrm{nm}$ \cite{Gaunt2013,Ville2017} (see Fig. \ref{fig:sketch}); a `tube' beam is created by a wide Gaussian beam ($125\mu\mathrm{m}$ waist radius) that has a circular hole at its center, created by a digital mirror device \cite{Hueck2017}.
The other two `end-cap' beams are created by two highly elliptical Gaussian beams with waist radii of $5.5\mu$m and $180\mu$m. Together, they generate a dark cylindrical volume with an approximate height of $39\mu$m and a diameter of $55\mu$m, defined by the full width at half maximum of the atomic density. The cylinder symmetry axis is parallel to the gravitational force. 

The experimental sequence starts by cooling the gas to quantum degeneracy in a crossed optical dipole trap \cite{Shkedrov2018}. To improve the loading efficiency into the flat trap, we have added a second crossing beam to the optical trap described in Ref. \cite{Shkedrov2018}. This yields a harmonic trap with trapping frequencies of $\omega_r= 2\pi \times 236(1)\mathrm{Hz}$ and $\omega_z= 2\pi \times 27(2)\mathrm{Hz}$, in the radial and axial directions, respectively. After forced evaporation, there are $N\approx 5\times10^5$ atoms at $T/T_F\approx 0.24$ in this trap, where $N$ is total atom number in both spin states and $T_{F}$ the Fermi temperature. 

\begin{figure}
	\centering
	\includegraphics{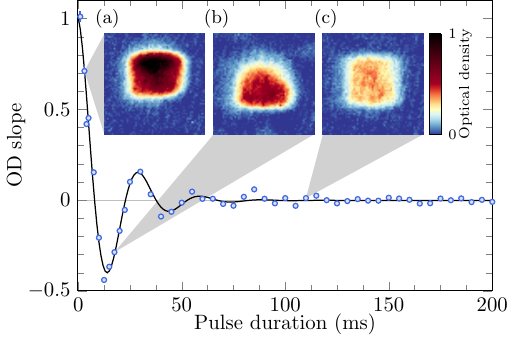}
	\caption{\textbf{Relaxation to a uniform density.} The gas is prepared in a spin-polarized $\left|\downarrow\right\rangle$ state without the rf field. The magnetic field is set according to Eq.(\ref{Eq:average_magnetic_gradient}), overcompensating gravity for this state by $5.9\%$. At $t=0$, the rf field with $\Omega/2\pi\approx15.7$kHz is switched on and flattens the time-average potential. The \emph{in situ} density of gas is recorded from the side of the cylindrical trap after variable rf pulse duration (insets showing $3\mathrm{ms}$, $17.5\mathrm{ms}$ and $110\mathrm{ms}$). The colorbar represents the optical depth (OD) in the absorption image. The main figure shows the density relaxation dynamics, quantified by the average change in the OD from side to side of the image. The errorbars represent $1\sigma$ deviation from linear fit to the OD slope. We fit the data with $g\left(t\right)=ae^{-t/\tau}\cos\left(2\pi ft+\phi\right)+b$ (solid line), and obtain $\tau=16(1)\mathrm{ms}$ and $f=33.6(7)\mathrm{Hz}$. The reduction of the OD for longer pulse duration is due to decoherence to a spin-balanced mixture while measuring atoms only in $\left|\downarrow\right\rangle$ state. (see main text). This measurement is done at a magnetic field of $\sim 203$G, where the scattering length between states $\left|\downarrow\right\rangle$ and $\left|\uparrow\right\rangle$ is $a\approx-1152a_0$ ($a_0$ is Bohr radius).}
	\label{fig:inhomogeneity_vs_pulse_duration}
\end{figure}

To load the flat trap, the tube beam is ramped to $30\mathrm{mW}$ already at the beginning of the evaporation in the harmonic trap. The magnetic field gradient that counteracts gravity is ramped to its final value, as given in Eq.(\ref{Eq:average_magnetic_gradient}), in $0.5$s, $1$s before the harmonic trap is turned off. In our system, we have an additional undesirable small magnetic gradient of $\frac{dB_z}{dy}\approx 0.68$G/cm in the transverse direction, which we compensate with another pair of coils. The sequence continues with a ramp up of the caps and tube beams power to $50\mathrm{mW}$ and $150\mathrm{mW}$, respectively. The two traps are held overlapping for $50$ms, and then the harmonic trap is ramped down in $200$ms. We typically load around $32\%$ of the atoms into the flat trap. Finally, the atoms are cooled in the flat trap by evaporation, forced by ramping down the power of the caps and tube beams in $2$s to $20\mathrm{mW}$ and $50\mathrm{mW}$, respectively. To ensure the cloud has reached equilibrium, we wait for an additional $0.8$s before performing a measurement. The final typical conditions in the flat trap with the rf field are $N\approx 60\times 10^{3}$ atoms with $T/T_{F}\approx 0.15$. The typical Fermi energy, $\epsilon_F/h\approx 940$Hz, is determined independently from an \emph{in situ} density measurement of the gas (see Appendix \ref{app:setting_EF}). The magnetic field is tuned around the Feshbach resonance, at $202.14$G, determining the strength of interactions, $1/k_Fa$. It is ramped adiabatically to its final value in $10$ms, where it is typically kept for $400\mathrm{ms}$. In the last $200\mathrm{ms}$ of the experimental sequence, the rf pulse is turned on with a typical Rabi frequency of around $\Omega/2\pi\approx10.5$kHz.

\section{Results}\label{Results_sec}
\textbf{\textbf{Relaxation dynamics.}---} Prior to turning the rf field on, the atomic densities of the two spins are not uniform (Fig.~\ref{fig:inhomogeneity_vs_pulse_duration}a), because the magnetic field gradient given by Eq.(\ref{Eq:average_magnetic_gradient}) over (under) compensates gravity for state $\left|\downarrow\right\rangle$ ($\left|\uparrow\right\rangle$) by $5.9\%$. Once the resonant rf field is turned on, the densities start to equilibrate. We study this relaxation process by preparing the gas with only spin $\left|\downarrow\right\rangle$ atoms, and imaging them from the side of the cylinder after different waiting times (Fig.~\ref{fig:inhomogeneity_vs_pulse_duration}a-c). To quantify the non-uniformity of the gas, we plot the slope of the optical depth (OD) at the center of the trap, normalized by its maximal value at $t=0$ (main part of Fig.~\ref{fig:inhomogeneity_vs_pulse_duration}). The gas relaxes to a uniform density with damped oscillations. The oscillation frequency is roughly given by the time it takes an atom with a Fermi velocity to traverse the trap height back and forth. The density reaches a steady-state for rf pulse duration longer than $100\mathrm{ms}$. The residual density inhomogeneity due to the finite steepness and imperfections of the trapping potential is discussed in Appendix \ref{app:setting_EF}.

Since spin-polarized fermions do not interact via s-wave scattering, a question that may arise is how the relaxation process actually occurs. While the rf rotation creates a $\left|\uparrow\right\rangle$ component on a short time-scale of the inverse Rabi frequency, the gas remains spin polarized, albeit with the spin oriented in a different direction. However, small spatial inhomogeneities in the rf and magnetic fields lead to spin dephasing, and eventually, through atomic diffusion, also to decoherence. Therefore, the spin-polarized gas becomes a balanced spin mixture on a relatively short timescale of $10$ms \cite{Gupta2003}. Thanks to the invariance of the interaction Hamiltonian, the rf resonance frequency does not change as the gas transforms from being non-interacting to strongly-interacting \cite{Zwierlein_2003}.

\textbf{\textbf{Momentum distribution of the uniform gas.}---} An important issue to consider is heating, which may occur during the initial relaxation phase and during the continuous operation of the rf pulse. We obtain the temperature of the gas by measuring its momentum distribution. Ordinarily, this is done by letting the gas expand ballistically either in free space or in a harmonic trap \cite{Tung2010}. Due to the relatively large initial size of the cloud, the free expansion requires particularly long expansion times, which are not always feasible. Expansion in a harmonic trap, on the other hand, is done for a quarter of the trap period, but is sensitive to anharmonicity of the trap \cite{Tung2010,Murthy2014,Mukherjee2017,Hueck2018}. 

\begin{figure}
\centering
\includegraphics{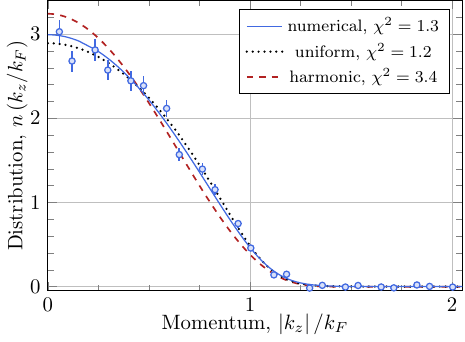}
\caption{\textbf{The one-dimensional momentum distribution of a uniform periodically-driven Fermi gas.} The distribution, measured by Raman spectroscopy \cite{Shkedrov2019}, is fitted with three models: a numerical model that provides a realistic description of the flat trap in the experiment (solid blue line, see Appendix \ref{app:setting_EF}), an ideal homogeneous Fermi-Dirac distribution given by Eq.(\ref{Eq:homogeneous_FD_mom_dist}) (dotted black line), and a harmonically trapped gas (dashed red line). The numerical and homogeneous models yield a better fit to the data than the harmonic one, based on the $\chi^2$ values. The differences between the uniform and numerical models are not significant. The temperature extracted from the numerical model is $T/T_F=0.15(1)$ with $T=7(2)\mathrm{nK}$. The results of the other fits are discussed in the main text. This measurement was performed at a magnetic field of $B=209.18$G, where the atoms are very weakly interacting. The rf pulse duration is $200$ms with a Rabi frequency of  $\Omega/2\pi\approx10.5$kHz. Error bars represent one standard deviation of the measured values.} 
\label{fig:raman_nk}
\end{figure}

Here, we take a different approach and use Raman spectroscopy, which has the advantage that it can be applied to a trapped gas \cite{Shkedrov2019,Ness2020}. The technique relies on a linear relation between the two-photon Raman detuning and the velocity of the atoms which are transferred from state $\left|\uparrow\right\rangle$ to the initially unoccupied state $|9/2,-5/2\rangle$. In the experiment, the two Raman beams are pulsed after the application of a $200$ms-long rf pulse. By scanning the relative frequency between the beams, we obtain a spectrum that is directly proportional to the one-dimensional momentum distribution \cite{Shkedrov2019}. A typical result with dynamically-driven uniform Fermi gas is shown in Fig.~\ref{fig:raman_nk}. The number of atoms in state $|9/2,-5/2\rangle$ is measured by selectively capturing them in a magneto-optical trap (MOT) and recording their fluorescence \cite{Shkedrov2018,Shkedrov2019}. To improve the detection, we separated the wavelength of the MOT, which is close to the D$_2$ transition, from that of the collected scattered photons \cite{Shkedrov2020a}. To this end, we added a dedicated probe beam, tuned to the D$_1$ transition, and filtered the recorded image with an ultra-narrow, $1$nm, optical band-pass filter \cite{Schlederer2021}. The intensities of the two Raman beams are actively stabilized and programmed to follow a $1$ms-long Blackman pulse \cite{Blackman1959}. The one-photon Raman detuning is around $46.1$GHz below the D$_1$ transition. To reduce unwanted single-photon scattering, which constitutes most of the background signal, we incorporated a temperature stabilized etalon after the Raman laser to filter the broadband amplified spontaneous emission.

We analyze the momentum distribution by fitting it with three different models (see Fig.~\ref{fig:raman_nk}). The first one is a numerical local-density approximation model of a gas in a realistic flat trap that is used in our experiment (solid blue line). In this model, we account for the finite steepness of the trap walls, which is calibrated using \textit{in situ} density images (see Appendix \ref{app:setting_EF}). The free parameters are the reduced temperature ($T/T_F$) and the background offset. For comparison, we fit the data with two ideal models of harmonically trapped gas (dashed red line) \cite{Shkedrov2019} and ideal uniform gas (dotted black line). The doubly-integrated momentum distribution of the latter is given by: 
\begin{equation}\label{Eq:homogeneous_FD_mom_dist}
   n\left(k_{z}\right)=\pi\frac{T}{T_{F}}\ln\left(1+\zeta e{}^{-\frac{k_{z}^{2}/k_{F}^{2}}{T/T_{F}}}\right) \ \ ,
\end{equation}
where $k_{z}$ is in the direction of the two-photon momentum transfer \cite{Shkedrov2019}, and $\zeta$ is the fugacity with the implicit form: $\mathrm{Li}_{\frac{3}{2}}(-\zeta)=-\frac{4}{3\sqrt{\pi}}\left(\frac{T}{T_{F}}\right)^{-3/2}$, with $\mathrm{Li}_n(z)$ being the Polylogarithm function. Notice that due to the double-integration, there is no sharp Fermi surface in this functional even at $T=0$.

The numerical and uniform models fit the data markedly better than the harmonic one, as evident by comparing their $\chi^2$ fit values (see legend of Fig.~\ref{fig:raman_nk}). The temperature extracted from the numerical model is $T/T_F=0.15(1)$ with $T=7(2)\mathrm{nK}$. The uniform model yields a very similar result of $T/T_{F}=0.16(2)$. In contrast, the harmonic model gives a much lower temperature of $T/T_{F}=0.08(2)$. In the numerical model, $k_F$ is calculated directly from \emph{in situ} density images (see Appendix \ref{app:setting_EF}), while in the uniform and harmonic models, it is left as a free fitting parameter. We find that the $k_F$ extracted from the uniform model is only $6(2)$\% higher than the one calculated directly, while that of the harmonically trapped model is higher by $30(2)$\%.

\begin{figure}[t]
\centering
\includegraphics{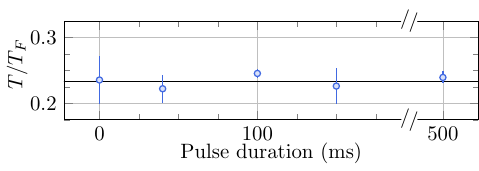}
\caption{\textbf{Reduced temperature as a function of the rf pulse duration.} The temperature is extracted from one-dimensional momentum distribution measurements fitted by a homogeneous Fermi-Dirac distribution. We observe no increase in $T/T_F$ up to $500\mathrm{ms}$, and this holds true also for $T$ alone. Note that the data in this figure were taken at slightly different conditions than the one in Fig.~\ref{fig:raman_nk}, with $N\approx131\times10^3$ atoms at $T/T_F\approx 0.23$. The measurement was performed at a magnetic field of $B=209.18$G, where the atoms are very weakly interacting. The Rabi frequency is $\Omega/2\pi\approx10.5$kHz. Error bars represent one standard error of the fit.} 
\label{fig:TTFvsDuration}
\end{figure}

To test whether the rf-induced spin rotation causes heating, we repeat the Raman measurement at different rf pulse durations, shown in Fig.~\ref{fig:TTFvsDuration}. Within the experimental accuracy, we do not observe an increase of the temperature. This result is corroborated by the measurements of the condensate fraction versus time, described below. We therefore conclude that with our experimental parameters, satisfying $\Omega\gg \epsilon_F/\hbar$, the heating rate is too small to be detected.

\textbf{\textbf{Pair condensation.}---} We now turn to probe the many-body properties of a dynamically driven Fermi gas. When a spin-balanced Fermi gas is cooled below the critical temperature, $T_c$, atoms with opposite spins pair and condense, forming a fermionic superfluid \cite{Greiner2003,Regal2004,Zwierlein2005,Zwerger2012}. The value of $T_c$ depends on the interaction strength. The survival of superfluidity is a stringent test of our Floquet engineering scheme, since this phase is extremely sensitive to heating and differential forces acting on the spins.

In these experiments, we cool the gas below the superfluid transition at unitarity ($1/k_Fa\approx0$). At the end of the cooling stage, the magnetic field is ramped in $10$ms from $203.5\mathrm{G}$ (weak interactions) to unitarity. There, it is held for $5$ms, during which the atoms pair-up and condense. The magnetic field gradient and rf pulse are present during the last $200$ms, long enough to ensure equilibrium. Since during this time the magnetic field is changing, we program the rf frequency to track the resonance transition. 

We characterize the survival of the superfluid phase by measuring the condensate fraction, using the pair-projection technique \cite{Regal2004,Zwierlein2004a}. To this end, the trap is abruptly turned off and at the same time the magnetic field is ramped rapidly ($40\mu$s) to the BEC side of the resonance ($199.8\mathrm{G}$) \cite{Regal2004}. This procedure projects the loosely-bound pairs onto tightly-bound molecules. We then let the gas expand for $24$ms and measure the distribution using absorption imaging. For the imaging to work, we dissociate the molecules by ramping back the magnetic field to unitarity just before taking the image. When the gas is superfluid, the recorded density distribution is bimodal, with condensed pairs appearing as a pronounced central peak (see Appendix \ref{app:CF}).

\begin{figure}
	\centering
\includegraphics{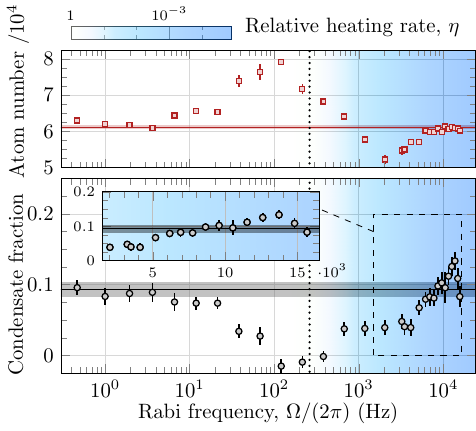}
    	\caption{\textbf{Condensate fraction and total atom number at unitarity as a function of the driving frequency.}
    	At very low frequencies, the spin oscillation is slow enough such that both observables are the same as without the driving, which are marked by horizontal lines with shading representing the uncertainty. At higher frequencies, the driving has an adverse effect on the superfluid. The vertical dashed line marks the frequency at which $\eta=1$  (see Eq.(\ref{eq_eta})). Above it, heating starts to be suppressed (the color represents the value of $\eta$). At even higher frequencies, the atom number returns to its initial value, and the condensate fraction reaches an even higher value than for a stationary gas (inset). The conditions are measured after $5$ms at the Feshbach resonance magnetic field, and the Rabi frequency is varied by changing the rf pulse power. The observables extraction procedure is discussed in Appendix~\ref{app:CF}. The error bars combine the $1\sigma$ mean confidence interval of the fit with statistical errors over $10$ repetitions.
    	}
	\label{fig:cf_vs_pulse}
\end{figure}

In Fig.~\ref{fig:cf_vs_pulse}, we plot the condensate fraction (lower panel) and total number of atoms (upper panel) at unitarity, extracted from the images of the expanded gas, as a function of the driving frequency. The overlap between the two spin distributions is large enough even without the rf field to yield a condensate fraction of around $0.11$. We distinguish between three frequency regimes with qualitatively different behavior. At very low frequencies, we observe a quasi-static behavior with local equilibrium and no apparent change in the conditions of the gas. As the frequency is increased, we cross to the second regime, where the driving generates spin currents and micro-motion that clearly harms the superfluid. Initially, at around $\Omega\sim 100$Hz, the heating does not lead to a loss of atoms because the energy of the excitation is smaller than the trap depth. In fact, as the condensate fraction decreases, the number of detected atoms increases. This surprising behavior is a result of a partial correlation between the number of detected atoms following the pair-projection technique and the number of pairs. It exists since during the magnetic field ramping a small fraction of the pairs are lost, most likely to deeply bound molecular states. At higher frequencies, atoms have enough energy to leave the trap, but heating is gradually suppressed due to the scaling with $\Omega$ of the second-order process (see section \ref{Theory_sec}). In the third regime, defined by $\eta<1$, the loss decreases and the condensate revives. In the high frequency limit, where $\eta\ll 1$, the atom number returns to its initial value while the condensate fraction reaches an even higher value, an effect we attribute to a better spatial overlap between the spins in a uniform gas.

\begin{figure}
	\centering
	\includegraphics{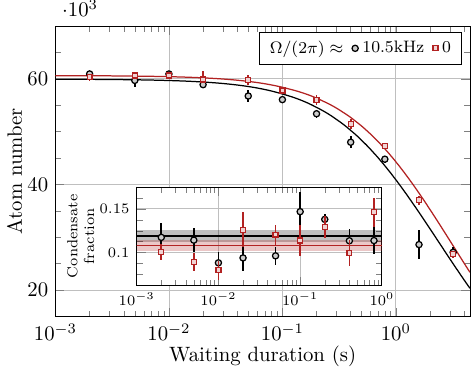}
    	\caption{\textbf{Time dependence of the atom number and condensate fraction at unitarity.} Data is taken after different waiting durations both with (black circles) and without (red squares) the rf driving. In both cases, the loss has a similar trend, which is well-fitted by the model in Eq.(\ref{Eq:loss_model}) (solid lines). Inset: The condensate fraction (same marks) is plotted together with the weighted average (solid lines), and its standard deviation (shades), shows no decrease. Error bars are determined as in Fig.~\ref{fig:cf_vs_pulse}.}
	\label{fig:cf_vs_waiting}
\end{figure}

We now return to the question of heating at high driving frequency. As shown in Fig.~\ref{fig:TTFvsDuration}, we do not observe a rise of the temperature, as extracted from the momentum distribution. However, in the superfluid phase, the condensate fraction is a much more sensitive thermometer. In  Fig.~\ref{fig:cf_vs_waiting}, we plot the total number of atoms and condensate fraction versus the waiting time at unitarity (black circles). We employ an rf field with a relatively high Rabi frequency ($\Omega/2\pi\approx10.5$kHz), where the density is already uniform and the condensate fraction reaches its high value (see Fig.~\ref{fig:cf_vs_pulse}). To distinguish between loss and heating due to the rf driving and other sources, we repeat this measurement without the rf field (red squares). The data, taken with a waiting time of up to more than $1$s, do not point to any heating, as there is no reduction of the condensate fraction, even when the total number of atoms has decreased by approximately a factor of two. 

The decay in the number of atoms is almost identical with or without the rf pulse.
We analyze the data using the following loss model \cite{Roberts2000,PhysRevLett.90.053201,Du2009} 
\begin{equation}\label{Eq:loss_model}
\frac{dn}{dt}=-K_{1}n-K_{2}n^{2}-K_{3}n^{3}\ \ ,    
\end{equation}
where $n$ is the total atomic density. $K_{1}=1/13.5\,\, \mathrm{ s^{-1}}$ is the single-body loss rate, determined by the rate of collisions with the residual gas in the vacuum chamber, and measured independently. $K_2$ and $K_3$ are the two- and three-body loss rate coefficients. Previously, these parameters were measured only with harmonically trapped gases, which complicated the analysis due to the non-linear density dependence in this model. Here, we benefit directly from the uniformity of the gas and from the fact its shape and volume are almost unchanged as the atom number diminishes. Fitting the data taken with the rf pulse with both coefficients as free parameters (black solid line) yields  $K_3=9(1)\times10^{-25}\mathrm{\ cm^6\  s^{-1}}$ and $K_2=0$.
This shows that the loss is mainly due to three-body recombination. Since in the data without the rf pulse the density is not homogeneous, and in fact, differs between the two spin components, we do not use it to extract loss coefficients. Qualitatively, however, it is still fitted well by the model of Eq.(\ref{Eq:loss_model}) (red solid line). Our value for $K_3$ is $10$ times higher than the one measured in a harmonic trap and at a significantly higher temperature in Ref.~\cite{PhysRevLett.90.053201}. We note, however, that their maximal value of $K_3$, which was measured on the BEC side of the resonance, agrees with our measurement at unitarity.

\textbf{\textbf{The contact parameter of a uniform gas.}---} We now turn to a measurement of the contact parameter in the BEC-BCS crossover regime. This parameter is central to a set of universal thermodynamic and energetic relations \cite{Tan08b,Tan08a,Tan08,Braaten08,Zhang2009,Braaten2012}, many of which have been tested experimentally \cite{PhysRevLett.104.235301,PhysRevLett.105.070402,PhysRevLett.95.020404,Castin2009}. Previous works determined the value of the contact with harmonically trapped gas at different temperatures and interaction strengths \cite{PhysRevLett.104.235301,Kuhnle2011a,Lingham2016}. Local measurements resolved the contact of a quasi-homogeneous sample \cite{Navon07052010,Sagi2012,Hoinka2013,Sagi2015,Horikoshi2017,Laurent2017,Carcy2019}. Until now, the contact of a truly uniform gas was measured only at unitarity \cite{Mukherjee2019}. 

\begin{figure}
\centering
\includegraphics{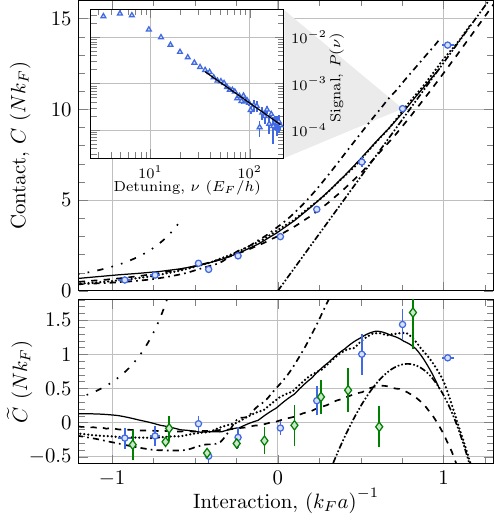}
\caption{\textbf{The contact of a uniform Fermi gas in the BEC-BCS crossover.} The contact is extracted from the tail of rf line-shapes taken at different interaction parameters. As an example, the inset shows the line-shape at $(k_Fa)^{-1}=0.75(2)$ together with its fit (black solid line). Upper panel: the theoretical prediction for the contact in the BCS (BEC) limit is shown as a loosely (densely) dash-double-dot line. We compare our data (blue circles) with a non-self-consistent T-matrix model at $T=T_{c}$ improved by Popov (dash-dotted line) and at $T=0$ (solid line) \cite{Palestini2010}, a Luttinger-Ward calculation (dashed line) \cite{Haussmann2009} and a GPF calculation (dotted line) \cite{Hu2006,Kuhnle2010}. The FNDMC line \cite{Gandolfi2011,Horikoshi2017} is indistinguishable from the GPF line on this scale. Lower panel: for a better quantitative comparison, we defined a shifted contact as $\widetilde{C}=C-3\exp\left(1.4/k_{F}a\right)$. In this plot we also add previous measurements done with a quasi-homogeneous gas (green diamonds) \cite{Sagi2015}. Error bars represent $1\sigma$ confidence interval of the fit.} 
\label{fig:Homogenous-contact-of}
\end{figure}

We determine the contact from the power-law tail of rf line-shapes taken with the uniform gas \cite{PhysRevLett.104.235301,Sagi2012,Shkedrov2018,Mukherjee2019}. In contrast to the condensate fraction experiments, where the condensate was formed after $\sim2$ms, here we observed that it takes at least $100$ms for the tail to fully develop. For this reason, we wait for $400$ms in the final magnetic field before measuring the rf line-shape. The number of atoms and temperature are similar to the condensate fraction experiments.
The spin-rotation rf field, with $\Omega/2\pi\approx10.5$kHz, is turned on for the last $200\mathrm{ms}$. It is turned off $0.5$ms before we probe with a $1\mathrm{ms}$ square pulse of a second rf field, whose frequency we scan near the $\left|\uparrow\right\rangle \rightarrow |9/2,-5/2\rangle$ transition. The atom number in state $|9/2,-5/2\rangle$ is again detected with fluorescence imaging \cite{Shkedrov2018}. A typical rf lineshape is shown on a logarithmic scale in the inset of Fig.~\ref{fig:Homogenous-contact-of}. A universal power-law tail over two decades is clearly visible. To extract the contact, we work in natural Fermi units and normalize the spectrum to ${1/2}$. The tail is then fitted with $C/(2^{2/3}\pi^2\nu^{3/2})$ (black line in the inset), where $C$ is the contact parameter in units of $Nk_F$. Owing to the high sensitivity of our fluorescence detection scheme, we keep the rf power constant for all detunings, while the maximal transferred fraction is no more than $8$\%. The systematic error in the determination of the contact due to the remaining density inhomogeneity in our realization of the flat trap is estimated by calculating the density-weighted average of a theoretical contact \cite{Palestini2010} using our calibrated model of the trap (see Appendix \ref{app:setting_EF}). We find that the density-averaged contact, representing our measured contact, is lower than the homogeneous contact by $5$\% in the BCS side ($1/k_Fa=-1$), and higher by $7$\% in the BEC side ($1/k_Fa=1$). Near unitarity, the large scattering length reduces the deviation to less than $1$\%. As a comparison, for harmonically trapped gas at the same average density, the systematic errors are $19$\% and $25$\% in the BCS and BEC sides, respectively.

In Fig.~\ref{fig:Homogenous-contact-of} we plot the contact of a uniform Fermi gas at various interaction strengths in the BEC-BCS crossover. Starting from the BCS side ($a<0$), the contact increases monotonically towards the BEC side of the crossover, where it converges to the asymptotic behavior of a molecule, $C_{\mathrm{BEC}}=4\pi/k_{F}a$ \cite{Haussmann2009}. We find that already above $1/k_{F}a\approx0.8$, our data are very close to $C_{\mathrm{BEC}}$. In contrast, the weak-coupling BCS limit of the contact, $C_{\mathrm{BCS}}=4\left(k_{F}a\right)^{2}/3$ \cite{Haussmann2009}, is not attained even at $1/k_{F}a=-1$. 
We compare our data to several theories and previous measurements. On the BEC side, there is a pronounced difference between the $T=0$ and $T=T_{c}$ predictions \cite{Palestini2010}. Our data, which were taken slightly below $T_{c}$, agrees with the $T=0$ T-matrix calculation. We also find a good agreement with the Gaussian pair fluctuations (GPF) calculation \cite{Hu2006,Kuhnle2010} and fixed-node diffusion Monte Carlo simulation (FNDMC) \cite{Gandolfi2011,Horikoshi2017}, especially in the BEC region. A Luttinger-Ward calculation \cite{Haussmann2009} is slightly below our data on the BEC side. Close to unitarity ($1/k_{F}a=-0.02(1)$), our measured $C=2.99(11)$ is in good agreement with values measured using rf spectroscopy, $C=3.07(6)$ \cite{Mukherjee2019}, Bragg spectroscopy $C=2.95(14)$ \cite{Carcy2019}, and impurity loss $C=3.09(34)$ \cite{Laurent2017}. It is slightly lower than $C=3.51(18)$\cite{Navon07052010} and $C=3.37(4)$\cite{Horikoshi2017} obtained from \emph{in situ} thermodynamic measurements. Similar data taken in the BEC-BCS crossover, albeit with a quasi-homogeneous gas and above $T_c$, is in agreement with ours, to within the experimental accuracy \cite{Sagi2015}.

\section{Discussion}\label{Discussion_sec}
In this work, we have demonstrated that Floquet engineering can be used with a continuous interacting Fermi gas without affecting its intrinsic many-body behavior. Specifically, we have employed the technique to eliminate the effect of a spin-dependent potential and achieve a flat trap. Our experiments are done with a driving frequency that is much higher than all other relevant experimental scales. In this regime, we have found no detectable heating during the experiment due to periodic driving. Measurements of the condensate fraction and contact parameter show the same behavior as expected in a stationary uniform Fermi gas. Furthermore, our dynamical levitation scheme can be used to generate a uniform density of other spin mixtures. 

The full Hamiltonian of Eq.(\ref{eq_Hamiltonian}) depends explicitly on time and therefore does not conserve energy. In contrast to many-body localized systems \cite{Ponte2015a,Lazarides2015,Khemani2016,Bordia2017}, our gas is ergodic and thus it is not protected from heating \cite{Lazarides2014,DAlessio2014,Ponte2015}. Nevertheless, interacting many-body systems can attain long lived prethermal states, following a quench of the Hamiltonian parameters \cite{Berges2004,Moeckel2008,Eckstein2009,Moeckel2010,Gring2012} or upon initiation of periodic driving \cite{Kuwahara2016,Abanin2017,Weidinger2017,Abanin2017a}. In particular discrete lattice systems show an exponentially slow energy absorption rate at high frequencies leading to Floquet prethermal states that persist for a time exponentially long with the drive frequency $\Omega$ \cite{Mori2016,Kuwahara2016,Abanin2017,Else2017,Machado2019,PhysRevX.10.021044}. 
The Fermi gas in the continuum does not benefit from such exponential suppression of the heating rate. Nonetheless, we have shown 
in section \ref{Theory_sec} that the smooth spatial structure of the periodic perturbation leads to parametric suppression of the heating rate in a unitary Fermi gas. The suppression is controlled by the small values of $V_g/(\hbar \Omega)$, with $V_g$ being the strength of the time-dependent perturbation, $\epsilon_F/(\hbar \Omega)$, and $1/{k_F L}$. This allows to obtain prethermal states with lifetimes much longer than the characteristic many-body timescale. In this work, we have focused on demonstrating this fact with measurements of the momentum distribution, the contact parameter, and the condensate fraction. It will be interesting to investigate the heating process and thermodynamic properties of the driven gas in the future.

Our work should also be placed in the framework of quantum information processing, where ultracold atoms are proposed as a resource for quantum memory \cite{Schnorrberger2009,Hsiao2018,Wang2019}.
A common cause of decoherence is spatially inhomogeneous spin-dependent potentials. As an example, the energy difference between two internal states of optically trapped atoms usually varies in position due to differential light shifts. In a classical ensemble, each atom can be treated independently with the rest of the ensemble acting as a fluctuating bath \cite{PhysRevLett.104.253003,PhysRevLett.105.093001}. These fluctuations lead to decoherence of the qubit stored in this atom. Dynamical decoupling \cite{PhysRevA.58.2733,PhysRevLett.82.2417}, a generalization of the celebrated Hahn echo technique \cite{PhysRev.80.580}, can substantially slow this relaxation process by applying multiple spin rotations \cite{PhysRevLett.105.053201}.

Dynamical decoupling has been applied successfully in NMR \cite{Carr1954,Meiboom1958,Haeberlen1976}, photonic systems \cite{Damodarakurup2009}, trapped ions \cite{Biercuk2009,Kotler2013}, electron spin in solids \cite{Du2009a,Lange2010,Naydenov2011,BarGill2013}, ultracold atoms \cite{PhysRevLett.105.053201,Almog2011}, and Bose-Einstein condensates (BEC) \cite{Trypogeorgos2018,Edri_2021}. In all cases, the decoupled system was weakly interacting and could be treated in a mean-field approach. Dynamical decoupling was not applied before to a strongly interacting ensemble with the aim of preserving its many-body behavior. In our work, the spin-dependent potential originates from the magnetic cancellation of gravity. The spin-rotation rf pulse we apply is a continuous version of a simple dynamical decoupling scheme. The absence of heating we observe is promising for future explorations of more sophisticated sequences tailored to generate specific local and global symmetries \cite{PhysRevLett.125.080602}. For example, the realization of tilted Fermi-Hubbard chains where the rf dressing was used to tune a relative tilt difference of two different spin states \cite{kohlert2021}.

\begin{acknowledgments}
We thank Amir Stern, Ari Turner, Netanel Lindner, Keiji Saito and David Huse for helpful comments. This research was supported by the Israel Science Foundation (ISF), grants No. 1779/19 and No. 218/19, and by the United States - Israel Binational Science Foundation (BSF), grant No. 2018264.
\end{acknowledgments}

\appendix

\section{Numerical model of the flat trap}\label{app:setting_EF}

We developed a systematic approach to calibrate a numerical model for the flat trap potential. Using the numerical model we calculate the Fermi energy and generate a momentum distribution functional used to fit the Raman spectrum (see Fig.~\ref{fig:raman_nk}). To this end, we simulate the density in the box trap using a model potential, and fit it to the \textit{in situ} integrated density, measured by absorption imaging. For this calibration, we create a spin-polarized gas at the same conditions as in the experiments presented in this paper. This is done by first preparing the gas in the flat trap in a spin-balanced configuration as described in the main text. Then we apply an adiabatic rapid passage selectively from state $\left|\uparrow\right\rangle$ that drives the atoms from this state to a final $|9/2,+9/2\rangle$ state, leaving state $\left|\downarrow\right\rangle$ untouched. The force created by the magnetic gradient, initially working opposite to the gravity for the $\left|\uparrow\right\rangle$ state, flips its sign due to the change in the magnetic number $m_F$ and starts working in the direction of gravity, ripping the atoms from the flat trap through the lower cap wall. This procedure removes all of the atoms that were initially in state $\left|\uparrow\right\rangle$ while loosing less than $10\%$ from state $\left|\downarrow\right\rangle$. The magnetic field gradient is set to perfectly cancel gravity for state $\left|\downarrow\right\rangle$, making the density homogeneous. 

\begin{figure}
\centering
\includegraphics{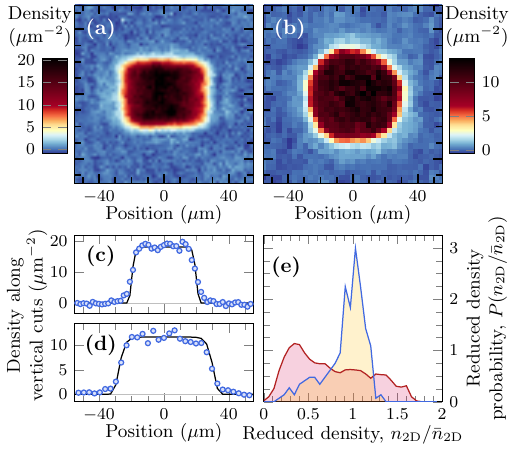}
\caption{\textbf{\emph{In situ} density analysis.} An example of an \emph{in situ} atomic density measured by absorption imaging from the side (a) and from the top (b) (the cylinder symmetry axis is vertical). A vertical cuts along the side and top images (circles) together with a fit to the numerical model (solid lines) are shown in insets (c) and (d), respectively. The flat part of the cut is where the density is uniform. Inset (e) shows a $2$D density probability distribution of the flat trap (blue line) and a harmonically trapped gas (red line). The density probability function of the flat trap is peaked around the average density value, indicating that most of the density is close to the average. Images (a) and (b) are an average of $5$ and $10$ experimental repetitions, respectively.}
\label{fig:In-situ-density from side}
\end{figure}

The next step is taking \textit{in situ} absorption images of the spin-polarized gas (see Fig. \ref{fig:In-situ-density from side}). The OD of the gas is too high to image directly. To reduce the OD for imaging, we apply a sequence of two rf pulses. The first pulse transfers $\approx 90\%$ of the atoms from state $\left|\downarrow\right\rangle$ to state $\left|\uparrow\right\rangle$. The second pulse transfers all of the atoms from state $\left|\uparrow\right\rangle$ to state $|9/2,-5/2\rangle$, which is detuned by $92\mathrm{MHz}$ from the optical transition. The last pulse ensures that no artifacts are introduced to the imaging due to large atom number in off-resonant states. The two pulses are completed within less than $70\mu$s, ensuring the density is unchanged during this procedure.

The spin-polarized gas is essentially non-interacting and can be described by a Fermi-Dirac distribution. To fit the two-dimensional integrated density image, we calculate the density in a local density approximation \cite{Ketterle2008}:
\begin{equation}
    n\left(r,z\right)=-\lambda_{dB}^{-3}\mathrm{Li}_{3/2}\left[-\exp\left(\beta\left[\mu-U\left(r,z\right)\right]\right)\right]\ \ ,
\end{equation}
where $\lambda_{dB}$ is the de Broglie wavelength, $\beta=1/(k_B T)$ with $k_B$ being the Boltzmann constant, $U\left(r,z\right)$ is the trapping potential and $\mu$ is the chemical potential, which is set by the total number of atoms. The model potential of the tube beam is parametrized by a power law function, while the potential of the cap beams is taken as a Gaussian function:
\begin{align}
&U\left(r,z\right)=U_{r}\left(r/\sigma_{r}\right)^{p}\\
&+U_{z}\left(\exp\left[-\frac{2\left(z-z_{0}\right)^{2}}{\sigma_{z}^{2}}\right]
+\exp\left[-\frac{2\left(z+z_{0}\right)^{2}}{\sigma_{z}^{2}}\right]\right)\nonumber \ \ .
\end{align}
Here $r$ is the radial coordinate relative to the symmetry axis of the tube (denoted by $z$), and $U_{r}$ is the potential barrier of the cylindrical wall. The tube radius $\sigma_{r}=32\mu$m and the power-law exponent $p=13.6$ are extracted from a direct measurement of the laser beam that generates the potential \cite{Shkedrov2020a}. $U_{z}$ is the potential barrier of the cap beams, $\sigma_{z}=7.5\mu$m is their waist radii in the z direction, also measured directly. $z_{0}=24\mu$m is half the separation between the two cap beams, measured by imprinting the caps profile on a dilute expanded cloud of atoms \cite{Shkedrov2020a}. The temperature, $T$, is found self-consistently together with momentum distribution measurements. The only free parameters in the fit are $U_{r}$, $U_{z}$ and the center position of the fit. A two-dimensional density function for fitting the \emph{in situ} image is generated by integrating the three-dimensional density $n\left(r,z\right)$ along one of the Cartesian axes. An example of a typical calibration is shown in Fig. \ref{fig:In-situ-density from side}. Once the model potential is calibrated, we calculate $\epsilon_F$ from the peak density, $E_{F}=\left(6\pi^{2}n_{F}\left(0,0\right)\right)^{2/3}\hbar^{2}/2m$. The density-averaged $\epsilon_F$ is smaller by $12$\%. This difference is considerably smaller than in the harmonically trapped gas before loading the flat trap, in which the density-averaged $\epsilon_F$ is smaller by $44$\% than the peak $\epsilon_F$.

We quantify the density homogeneity in the form of a probability distribution function $P\left(n\right)$. We calculate the axially-integrated probability distribution $P\left(n_{2D}/\bar{n}_{2D}\right)$ as an histogram of absorption images taken along the symmetry axis (inset (b) of Fig. \ref{fig:In-situ-density from side}), where $n_{2D}$ is the two-dimensional atomic density and $\bar{n}_{2D}$ is its weighted average. The probability distribution is plotted in inset (e) of Fig. \ref{fig:In-situ-density from side}, together with the probability distribution of a harmonically trapped gas. A direct measure of uniformity is the standard deviation of $P\left(n_{2D}/\bar{n}_{2D}\right)$, which is 0.23 and 0.42 in units of $\bar{n}_{2D}$ for the flat and harmonic traps, respectively. 

The knowledge of the trapping potential enables us to calculate the one-dimensional momentum distribution $n\left(k_{z}\right)$ and use it as a fitting function for the data acquired in the Raman spectroscopy experiment (Fig.~\ref{fig:raman_nk}). $n\left(k_{z}\right)$ is obtained by integrating the semi-classical distribution \cite{Ketterle2008}
\begin{equation}
    f\left(\boldsymbol{r},\boldsymbol{k}\right)=\left(\exp\left[\left(\frac{\boldsymbol{k}^{2}}{k_{F}^{2}}+\frac{U\left(r,z\right)}{E_{F}}-\frac{\mu}{E_{F}}\right)/\frac{T}{T_{F}}\right]+1\right)^{-1}
\end{equation} 
both spatially and along two momenta axes 
\begin{equation}
    n\left(k_{z}\right)=\frac{4\pi}{3N}\frac{1}{\left(2\pi\right)^{3}}\int f\left(\boldsymbol{r},\boldsymbol{k}\right)d^{3}\boldsymbol{r}dk_{x}dk_{y} \ \ .
\end{equation}

\section{Extracting the condensate fraction}\label{app:CF}
 
\begin{figure}
\centering
\includegraphics{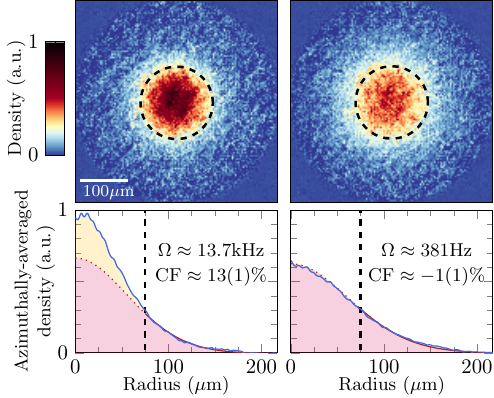}
\caption{\textbf{Extraction of the condensate fraction.} Upper panels are absorption images of high (left) and low (right) condensate fraction averaged over $10$ experimental repetitions. Lower panels present the corresponding azimuthally-averaged signals (blue lines). We fit Eq.(\ref{eq:g_1_2_n_k}) (red line) to the thermal wings at radii $>R_\mathrm{mask}$ (dashed black line). The condensate fraction is defined by the integrated signal above the fit extrapolation (yellow shading) over the total integrated signal (pink plus yellow shadings). To make a fair comparison, each of the two distributions is normalized by its total number of atoms, $61.0\times10^3$ and $68.3\times10^3$ in the right and left examples, respectively.
}
\label{fig:CF_fit}
\end{figure}

To separate between the thermal wings and the central peak, we image the cloud after a relatively long expansion. As a result, the absorption signal is weak. To improve the signal-to-noise ratio, we employ a deep-learning approach to filter out the background noise in the images \cite{Ness2020a}. We have verified that this noise removal procedure does not change significantly the reported CF values and only reduces the uncertainty.
The recorded density can be roughly considered as dissociated pairs that have either non-zero or zero center-of-mass momentum. The latter are the condensed pairs which constitute the central peak of the image (see Fig.~\ref{fig:CF_fit}). Each of the two populations expands differently, according to their respective momentum distribution. The total atom number is extracted by a direct integration of the OD image, and the error bars indicate statistical standard error only.

Following the time of flight, the non-condensed part of the gas is characterized by a wider expansion with respect to the trap dimensions, while the condensed part spreads only slightly beyond its original size, set by the tube beam diameter \cite{Mukherjee2017}. To separate between the condensed and non-condensed parts, we mask out the central region of the image (dashed line in Fig. \ref{fig:CF_fit}) and fit only the tail of the azimuthally-averaged signal with the momentum distribution of a thermal gas of non-interacting bosons \cite{Ketterle2008} (pink shading in Fig.~\ref{fig:CF_fit}),
\begin{equation}
    \label{eq:g_1_2_n_k}
    n_{\mathrm{th}}(k)=\frac{V}{4\pi^2\lambda_T}\mathrm{Li}_{1/2}\left(z e^{-\lambda_T^2 k^2/4\pi}\right)+N_{\mathrm{bg}}\;,
\end{equation}
where the thermal de Broglie wavelength $\lambda_T=h/\sqrt{2\pi mk_B T}$ and the fugacity $z$, are fitted under the normalization constraint $N_{\mathrm{th}}=2\pi\int n_{\mathrm{th}}(k)kdk$, and $N_{\mathrm{bg}}$ accounts for the background signal.
We extract the condensed population from the signal that lies above the fit (yellow shading in Fig.~\ref{fig:CF_fit}). The mask radius, $R_\mathrm{mask}$, should be large enough to leave only the thermal wings for fitting. When we analyze data taken with no condensate, we observe that the width of the fitted distribution is almost independent of the mask radius up to around $R_\mathrm{mask}\approx 80\mu$m. For larger radii, the signal in the remaining thermal wings is too weak, and the fit exhibits a systematic deviation. Therefore, we set the mask radius to $R_\mathrm{mask}=75\mu$m.  Finally, we note that the CF values we obtain are close to those reported in Ref.~\cite{Mukherjee2019}, taking into account our measured reduced temperature.

\newpage

%

\end{document}